\documentclass[%
 reprint,
superscriptaddress,
 amsmath,amssymb,
 aps,
prl,
]{revtex4-2}
   
\usepackage{graphicx}% Include figure files
\usepackage{dcolumn}% Align table columns on decimal point
\usepackage{bm}% bold math
\usepackage[caption=false]{subfig}
\usepackage{xcolor}
\usepackage{physics}
\usepackage{comment}
\usepackage{amsmath}
\usepackage{tikz}
\usepackage[colorlinks=true,citecolor=blue,linkcolor=blue,urlcolor=blue]{hyperref}
\usetikzlibrary{arrows.meta} 
\usetikzlibrary{calc}

\begin{document}

\preprint{APS/123-QED}

\title{Controlling quantum transport by measurement-rate modulation}

\author{Jes\'{u}s Casado-Pascual}
\email{jcasado@us.es}
\affiliation{F\'{i}sica Te\'{o}rica, Universidad de Sevilla, Apartado de Correos 1065, 41080 Sevilla, Spain}%
\affiliation{Multidisciplinary Unit for Energy Science, Universidad de Sevilla, E-41080 Sevilla, Spain}%
\author{Luis Octavio Casta\~{n}os-Cervantes}
\email{luis.castanos@tec.mx}
\affiliation{Tecnol\'ogico de Monterrey, School of Engineering and Sciences, 14380, Mexico City, Mexico}%
\affiliation{Universidad Nacional Aut\'{o}noma de M\'{e}xico, Facultad de Ingenier\'{i}a, Av. Universidad 3000, Ciudad Universitaria, Coyoac\'{a}n 04510, Ciudad de M\'{e}xico, M\'{e}xico}

\date{\today}

\begin{abstract}
	Temporal modulation of the measurement rate provides a powerful mechanism for controlling open-system dynamics through measurement backaction. We demonstrate this mechanism in a minimal exactly solvable model of a continuously monitored quantum particle on a switching lattice. We derive exact analytical expressions for the long-time current under both periodic and stochastic switching, revealing a common slow-switching limit and identifying a measurement-induced crossover between Zeno- and anti-Zeno-dominated transport regimes that controls the direction of the current.
\end{abstract}

\maketitle

\paragraph*{Introduction---\hspace{-10pt}} Continuous monitoring provides a fundamental framework for understanding and manipulating quantum dynamics through measurement backaction~\cite{Wiseman2009}. Among its most prominent manifestations are the quantum Zeno effect~\cite{Misra1977,Itano1990,Facchi2008} and the rich phenomenology recently uncovered in monitored many-body systems~\cite{Li2018,Skinner2019,Chan2019}. More recently, measurement-induced motion has been shown to emerge from the symmetry properties of the monitored observables~\cite{Antonic2025}. Continuous monitoring is characterized by two fundamental ingredients: the observable being monitored and the measurement rate at which information is extracted from the system. Recent work has shown that the measurement rate can strongly influence transport in continuously monitored quantum systems~\cite{Ferreira2024}. Here we show that its temporal modulation provides a distinct mechanism for controlling stationary transport exclusively through measurement backaction. Specifically, we demonstrate that suitably designed continuous-monitoring protocols, in which the measurement rate is modulated in time, allow stationary properties to be engineered without modifying either the Hamiltonian, the dissipative environment, or the measured observable.

To gain physical insight into this mechanism, we introduce a minimal exactly solvable model describing a continuously monitored quantum particle moving on a one-dimensional lattice that switches between two configurations. In the absence of measurement-rate modulation, the transport tendencies associated with the two configurations exactly cancel each other, yielding zero stationary current. We show that assigning different continuous measurement rates to the two configurations breaks this dynamical balance, giving rise to a finite stationary particle current. The exact solvability of the model enables the underlying mechanism to be characterized analytically over the entire parameter space, revealing current reversals together with a nontrivial interplay between coherent evolution, incoherent transitions, and quantum-Zeno physics. More generally, it provides analytical insight into how temporal modulation of the measurement rate can be exploited to shape stationary properties of open quantum systems.

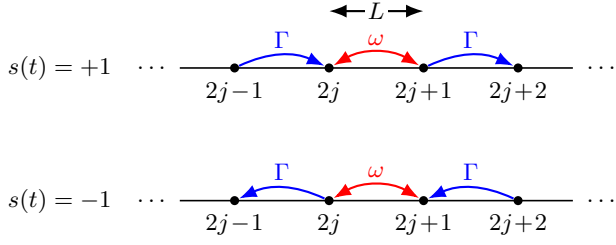
\begin{figure}[t]
	\centering

\begin{tikzpicture}[scale=1.25, >=Latex]
	
	\usetikzlibrary{arrows.meta}
	
	\tikzset{
		site/.style={circle, fill=black, inner sep=1.2pt},
		coh/.style={<->, thick, red},
		inc/.style={->, thick, blue},
		lab/.style={font=\small}
	}
	
	% Parameters
	\def\xline{2.08}
	\def\xdots{2.38}
	\def\ysep{-1.40}
	
	%==================== PANEL s(t)=+1 ====================
	
	\node at (-3.35,0) {$s(t)=+1$};
	
	\node at (-\xdots,0) {$\ldots$};
	\node at ( \xdots,0) {$\ldots$};
	
	\node[site,label=below:$2j\!-\!1$] (a1) at (-1.5,0) {};
	\node[site,label=below:$2j$]       (a2) at (-0.5,0) {};
	\node[site,label=below:$2j\!+\!1$] (a3) at ( 0.5,0) {};
	\node[site,label=below:$2j\!+\!2$] (a4) at ( 1.5,0) {};
	
	% lattice
	\draw[semithick] (-\xline,0) -- (\xline,0);
	
	% coherent tunneling
	\draw[coh,bend left=30] (a2) to (a3);
	
	% incoherent hopping
	\draw[inc,bend left=25] (a1) to (a2);
	\draw[inc,bend left=25] (a3) to (a4);
	
	% labels
	\node[lab,red]  at (0,0.30) {$\omega$};
	\node[lab,blue] at (-1,0.30) {$\Gamma$};
	\node[lab,blue] at ( 1,0.30) {$\Gamma$};
	
	% lattice spacing
	\draw[<->,thick]
	(-0.5,0.6)
	--
	node[fill=white,inner sep=1pt] {$L$}
	(0.5,0.6);
	
	%==================== PANEL s(t)=-1 ====================
	
	\node at (-3.35,\ysep) {$s(t)=-1$};
	
	\node at (-\xdots,\ysep) {$\ldots$};
	\node at ( \xdots,\ysep) {$\ldots$};
	
	\node[site,label=below:$2j\!-\!1$] (b1) at (-1.5,\ysep) {};
	\node[site,label=below:$2j$]       (b2) at (-0.5,\ysep) {};
	\node[site,label=below:$2j\!+\!1$] (b3) at ( 0.5,\ysep) {};
	\node[site,label=below:$2j\!+\!2$] (b4) at ( 1.5,\ysep) {};
	
	% lattice
	\draw[semithick] (-\xline,\ysep) -- (\xline,\ysep);
	
	% coherent tunneling
	\draw[coh,bend left=30] (b2) to (b3);
	
	% incoherent hopping (reversed)
	\draw[inc,bend right=25] (b2) to (b1);
	\draw[inc,bend right=25] (b4) to (b3);
	
	% labels
	\node[lab,red]  at (0,\ysep+0.30) {$\omega$};
	\node[lab,blue] at (-1,\ysep+0.30) {$\Gamma$};
	\node[lab,blue] at ( 1,\ysep+0.30) {$\Gamma$};
	
\end{tikzpicture}

\caption{
		Schematic representation of the lattice for the two possible values of the time-dependent lattice configuration $s(t)$. The lattice consists of equally spaced sites separated by a distance $L$. Coherent tunneling (red double-headed arrows) couples the two sites within each dimer $(2j,2j+1)$ with tunneling rate $\omega$, whereas incoherent hopping (blue arrows) connects neighboring dimers with transition rate $\Gamma$. For $s(t)=+1$, hopping proceeds from site $2j-1$ to site $2j$, whereas for $s(t)=-1$ the hopping direction is reversed. The two configurations therefore favor transport in opposite directions.
}
\label{fig:lattice}

\end{figure}

\paragraph*{Model---\hspace{-10pt}} We consider a single quantum particle moving on a one-dimensional tight-binding lattice consisting of equally spaced sites separated by a distance $L$ [Fig.~\ref{fig:lattice}]. The particle Hilbert space is spanned by the localized Wannier states $\{\ket{j}:j\in\mathbb{Z}\}$, where $\ket{j}$ denotes the state localized at lattice site $j$. The lattice switches between two configurations, labeled by a classical variable $s(t)\in\{+1,-1\}$. Two switching protocols are investigated. In the first, $s(t)$ alternates periodically between the two configurations with period $T$ according to
\begin{equation}
	s(t)=
	\begin{cases}
		+1, & t\in[0,T/2),\\
		-1, & t\in[T/2,T),
	\end{cases}
	\label{speriodic}
\end{equation}
where the pattern is repeated periodically for all subsequent times. In the second protocol, $s(t)$ follows a dichotomous Markov process that switches between the two configurations with rate $\gamma$, corresponding to a mean residence time $\gamma^{-1}$ in each configuration. Both protocols are considered to demonstrate the robustness of the proposed mechanism under deterministic and stochastic switching.

Conditioned on the instantaneous lattice configuration $s(t)$, the evolution of the particle density operator $\rho(t)$ is governed by the master equation
\begin{equation}
	\dot{\rho}(t)
	=
	-\frac{i}{\hbar}[H,\rho(t)]
	+\Gamma \,\mathcal{D}^{\mathrm{inc}}_{s(t)}[\rho(t)]
	+R_{s(t)}\,\mathcal{D}^{\mathrm{meas}}[\rho(t)],
	\label{eq:master}
\end{equation}
where the overdot denotes the time derivative. The three contributions on the right-hand side represent coherent tunneling, incoherent nearest-neighbor transitions, and continuous position measurements, respectively. We now discuss each contribution in turn.

The coherent dynamics is described by the Hamiltonian
\begin{equation}
	H=\hbar\omega\sum_{j\in\mathbb{Z}}
	\left(
	\ketbra{2j}{2j+1}
	+
	\ketbra{2j+1}{2j}
	\right),
	\label{eq:H}
\end{equation}
where $\omega$ is the tunneling frequency.  The Hamiltonian couples only the two sites within each dimer $(2j,2j+1)$, leaving different dimers completely decoupled.

Neighboring dimers are connected through incoherent hopping processes occurring at rate $\Gamma$. Their direction depends on the instantaneous lattice configuration $s(t)$ and is described by the jump operators
\begin{equation}
	L^{(j)}_{s(t)}
	=
	\delta_{s(t),+1}\, \ketbra{2j}{2j-1}
	+
	\delta_{s(t),-1}\,\ketbra{2j-1}{2j},
\end{equation}
which generate the dissipative contribution appearing as the second term on the right-hand side of Eq.~(\ref{eq:master}),
\begin{equation}
	\mathcal{D}^{\mathrm{inc}}_{s(t)}[\rho(t)]
	=
	\sum_{j\in\mathbb{Z}}
	\left[
	L^{(j)}_{s(t)}
	\rho(t)
	L^{(j)\dagger}_{s(t)}
	-
	\frac12
	\left\{
	L^{(j)\dagger}_{s(t)}
	L^{(j)}_{s(t)},
	\rho(t)
	\right\}
	\right].
\end{equation}
For $s(t)=+1$, the particle undergoes incoherent hopping from site $2j-1$ to the neighboring site $2j$, whereas for $s(t)=-1$ the hopping direction is reversed. Together with the coherent tunneling within each dimer, these configuration-dependent hopping processes define the hybrid coherent--dissipative dynamics shown schematically in Fig.~\ref{fig:lattice}.

The third contribution on the right-hand side of Eq.~(\ref{eq:master}) accounts for continuous monitoring of the particle position. The monitored observable is the particle position, corresponding to the projective measurement operators $\{\ketbra{j}{j}:j\in\mathbb{Z}\}$. Within the framework of continuous monitoring, these measurements give rise to a measurement-induced dephasing described by the dissipator
\begin{equation}
	\mathcal{D}^{\mathrm{meas}}[\rho(t)]
	=
	\sum_{j\in\mathbb{Z}} \ev{\rho(t)}{j}\ketbra{j}{j}
	-
	\rho(t).
\end{equation}
The measured observable is identical in both lattice configurations. The only configuration dependence of the monitoring process enters through the measurement rate,
$
	R_{s(t)}
	=
	\delta_{s(t),+1}R_+
	+
	\delta_{s(t),-1}R_-
$, 
where $R_+$ and $R_-$ denote the measurement rates associated with the lattice configurations $s(t)=+1$ and $s(t)=-1$, respectively.  Hence, the measurement rate is modulated in phase with the lattice switching, whereas the measured observable itself remains unchanged. The limit $R_+=R_-$ corresponds to uniform continuous monitoring, i.e., to the absence of measurement-rate modulation.

Although the present model is intended as a minimal theoretical description designed to identify the key ingredients of the proposed mechanism, experimental realizations may be envisaged using ultracold atoms trapped in programmable optical potentials. Remarkable progress has been made in the independent control of lattice geometries and tunneling amplitudes using programmable optical potentials~\cite{Henderson2009,Gaunt2012,Windpassinger2013}, in the engineering of dissipative quantum dynamics~\cite{Diehl2008,Verstraete2009}, in site-resolved position measurements by means of quantum-gas microscopy~\cite{Bakr2009,Sherson2010}, and in the observation of measurement backaction in optical lattices~\cite{Patil2015}. More recently, dissipative control has also been exploited to realize nonreciprocal quantum transport in ultracold-atom systems~\cite{Gou2020}. Taken together, these developments suggest that effective models combining the ingredients considered here may become experimentally accessible as programmable quantum platforms continue to advance.

\paragraph*{Reduced transport dynamics---\hspace{-10pt}} The transport properties of the system are characterized by the average particle velocity, as is customary in studies of directed transport~\cite{ReimannR,Hanggi2009,Salger2009,Dupont2023,CastanosCervantes2024}. Defining the position operator as
$
X=L\sum_{j\in\mathbb Z}j\,\ketbra{j}{j},
$
the instantaneous velocity is given by
$
v(t)=\frac{d}{dt}\ev{X}_t
=\Tr\!\left[X\dot{\rho}(t)\right].
$
The corresponding long-time averaged velocity,
\begin{equation}
	v_\infty=
	\lim_{\tau\to\infty}
	\frac1\tau
	\int_0^\tau v(t)\,dt
	=
	\lim_{\tau\to\infty}
	\frac{\ev{X}_{\tau}}{\tau},
	\label{LTZAV}
\end{equation}
characterizes the net directed transport.

Despite the infinite-dimensional Hilbert space, the transport dynamics admits an exact reduction to two collective dynamical variables. Using Eq.~(\ref{eq:master}), we find
\begin{equation}
	v(t)
	=
	\frac{L}{2}
	\left[
	2\omega y(t)
	-
	\Gamma\bigl(z(t)-s(t)\bigr)
	\right],
	\label{avvelo}
\end{equation}
where
$
y(t)=2\sum_{j\in\mathbb Z}
\Im\!\left[\rho_{2j,2j+1}(t)\right]
$
represents the total imaginary part of the intradimer coherences, whereas
$
z(t)=
\sum_{j\in\mathbb Z}
\left[
\rho_{2j,2j}(t)
-
\rho_{2j+1,2j+1}(t)
\right]
$
denotes the population imbalance between even and odd lattice sites. Here,
$\rho_{j,k}(t)=\mel{j}{\rho(t)}{k}$.

Differentiating the definitions of $y(t)$ and $z(t)$ with respect to time and using Eq.~(\ref{eq:master}), one obtains
\begin{align}
	\dot{y}(t)
	&=
	-\frac{1}{2}
	\left[
	\Gamma+2R_{s(t)}
	\right]
	y(t)
	+
	2\omega z(t),
	\label{ydot}\\
	\dot{z}(t)
	&=
	-2\omega y(t)
	-
	\Gamma\left[z(t)-s(t)\right].
	\label{zdot}
\end{align}
Hence, the infinite-dimensional transport problem admits an exact reduction to Eqs.~(\ref{ydot}) and (\ref{zdot}). Depending on the switching protocol, these equations constitute a system of linear differential equations with either piecewise constant coefficients or coefficients governed by a dichotomous Markov process. In both cases, exact analytical expressions for the long-time averaged velocity can be derived.

\paragraph*{Exact analytical solution---\hspace{-10pt}} The reduced equations derived above admit exact analytical solutions for both periodic and stochastic switching protocols, yielding explicit expressions for the long-time averaged velocity. Only the essential steps are presented below, while the complete derivations are given in Appendices~\ref{AppA} and \ref{AppB}.

For periodic switching, Eqs.~(\ref{ydot}) and (\ref{zdot}) are solved separately within each half-period, where the coefficients are constant, and the corresponding solutions are matched continuously at the switching times. The resulting dynamics converges to a unique asymptotic periodic state, described by $y_{\mathrm p}(t)$ and $z_{\mathrm p}(t)$, independently of the initial conditions. Evaluating Eq.~(\ref{avvelo}) with this asymptotic solution and averaging over one period yields the exact long-time averaged velocity
\begin{equation}
	v_{\infty}
	=
	\frac{
		16\kappa L\omega^2\Gamma^2\delta R
	}{
		\left[
		8\omega^2+\Gamma\left(\Gamma+2R_{+}\right)
		\right]
		\left[
		8\omega^2+\Gamma\left(\Gamma+2R_{-}\right)
		\right]
	},
	\label{velocity1}
\end{equation}
where $
\delta R=(R_{+}-R_{-})/2
$ and
\begin{equation}
	\kappa=
	\frac{\Delta y_{\mathrm p}}{\omega T}
	+
	\frac{2\Delta z_{\mathrm p}}{\Gamma T}
	-1.
	\label{kappa}
\end{equation}
Here,
$
\Delta y_{\mathrm p}=y_{\mathrm p}(T/2)-y_{\mathrm p}(0),
$
$
\Delta z_{\mathrm p}=z_{\mathrm p}(T/2)-z_{\mathrm p}(0)$.
Equation~(\ref{velocity1}) shows that directed transport is possible only when the measurement rate is modulated, i.e., for $\delta R\neq0$.
%Equation~(\ref{velocity1}) shows that directed transport requires different measurement rates, i.e., $\delta R\neq0$.

For stochastic switching, $s(t)$ is a dichotomous Markov process with transition rate $\gamma$. Since the switching process is ergodic and the dynamics is dissipative, the Birkhoff ergodic theorem~\cite{birkhoff1931} allows the long-time average entering Eq.~(\ref{avvelo}) to be replaced by the corresponding stationary ensemble average. Taking stationary averages of Eqs.~(\ref{ydot}) and (\ref{zdot}) yields a set of equations involving the moments $\langle y\rangle_{\mathrm{st}}$, $\langle z\rangle_{\mathrm{st}}$, $\langle sy\rangle_{\mathrm{st}}$, and $\langle sz\rangle_{\mathrm{st}}$. The mixed moments are evaluated exactly by means of the Shapiro--Loginov formula~\cite{shapiro1978,bena2006}, leading to
\begin{equation}
	v_{\infty}
	=
	-
	\frac{
		4L\omega^{2}\Gamma^{2}\delta R
	}{
		(\Gamma+2\gamma)
		\left[
		\Omega_{1}\Omega_{2}^{2}
		-
		\Gamma(\delta R)^{2}
		\right]
	},
	\label{v_final_random}
\end{equation}
where
$
\Omega_{1}=\Gamma/2+\bar{R}+2\gamma+4\omega^{2}/(\Gamma+2\gamma),
$
$
\Omega_{2}=\sqrt{\Gamma(\Gamma/2+\bar{R})+4\omega^{2}},
$
and
$
\bar{R}=(R_{+}+R_{-})/2.
$
Equation~(\ref{v_final_random}) likewise shows that directed transport requires measurement-rate modulation, i.e., $\delta R\neq0$.
%Equation~(\ref{v_final_random}) likewise shows that directed transport requires different measurement rates, i.e., $\delta R\neq0$.

\paragraph*{Physical interpretation---\hspace{-10pt}} The exact analytical expressions obtained above share a common asymptotic limit corresponding to slow switching, where the residence time in each lattice configuration greatly exceeds the intrinsic relaxation time of the system. This regime is reached as $T\to\infty$ ($\kappa\to-1$) for periodic switching and as $\gamma\to0^+$ for stochastic switching. In both cases,
\begin{equation}
	v_\infty
	\sim
	\frac{v_\infty^{(+)}+v_\infty^{(-)}}{2},
	\label{eq:commonlimit}
\end{equation}
where
$
v_\infty^{(\pm)}
=
\pm
8L\omega^2\Gamma/
\left[8\omega^2+\Gamma(\Gamma+2R_\pm)\right]
$
are the stationary transport velocities associated with the static lattice configurations $s=\pm1$. This result reflects the fact that, in the slow-switching regime, the system relaxes close to the stationary state corresponding to each configuration before the next switching event occurs, so that the asymptotic current is simply the arithmetic mean of the stationary currents associated with the two static configurations.

Equation~(\ref{eq:commonlimit}) also reveals the physical origin of the directed transport. In the slow-switching regime, each lattice configuration drives the particle in the direction of its corresponding stationary current. Since the two configurations are related by spatial inversion, they generate stationary currents of equal magnitude and opposite sign whenever $R_+=R_-$, resulting in zero net transport. If the measurement rates are different, however, the coherent dynamics is more strongly suppressed in the configuration with the larger measurement rate owing to the quantum Zeno effect. Consequently, the magnitude of the corresponding stationary current is reduced, breaking the exact cancellation between the two stationary currents. The stationary current associated with the smaller measurement rate therefore prevails, yielding $v_\infty>0$ for $R_->R_+$ and $v_\infty<0$ for $R_+>R_-$.

\begin{figure}[t]
	\centering
	 \includegraphics[width=\linewidth]{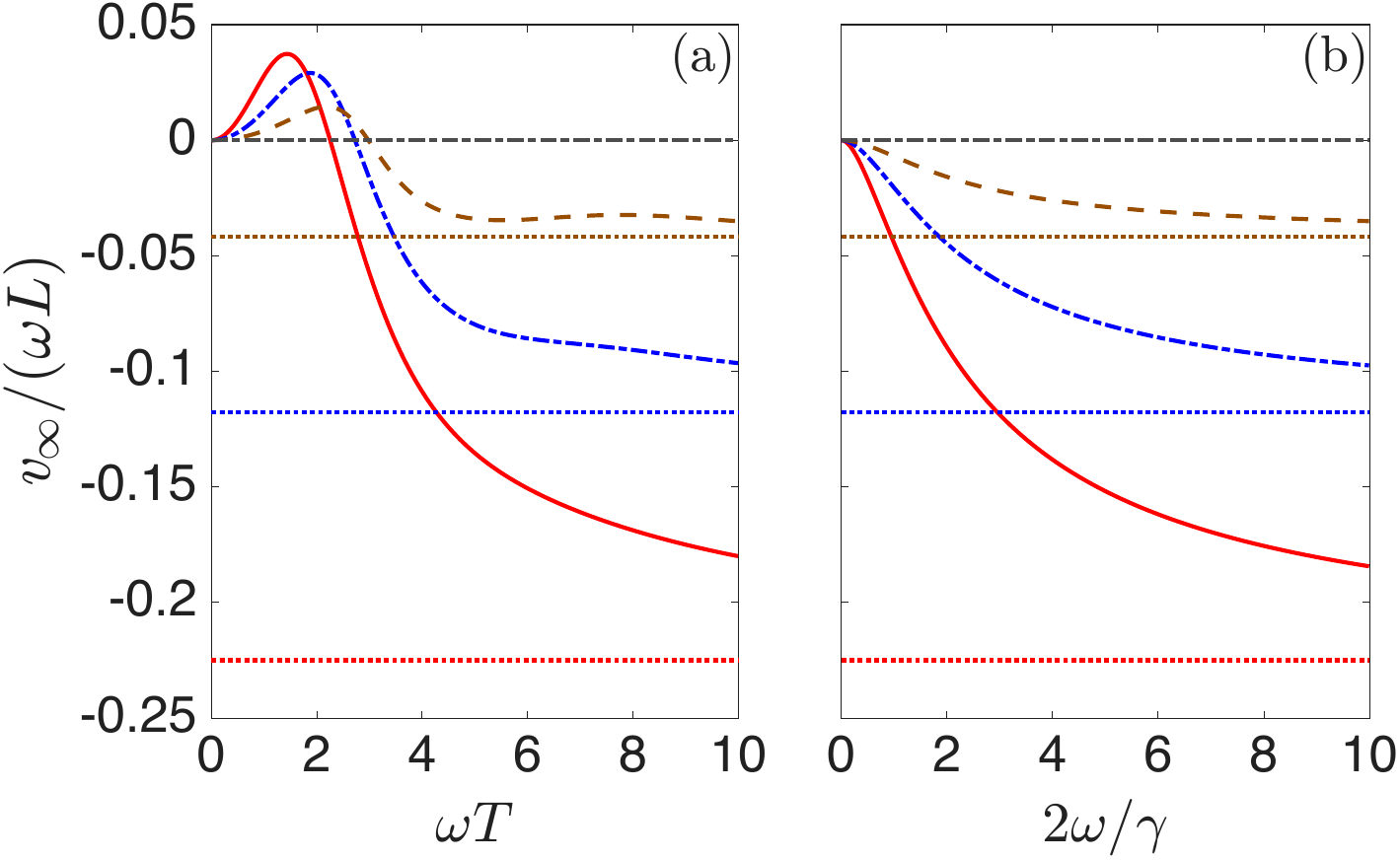}
	\caption{Dimensionless long-time averaged velocity $v_{\infty}/(\omega L)$ as a function of (a) $\omega T$ for the periodic protocol and (b) $2\omega/\gamma$ for the stochastic protocol. Whereas the stochastic protocol preserves the transport direction for all switching rates, the periodic protocol exhibits a current reversal as the switching period decreases, signaling a crossover from Zeno- to anti-Zeno-dominated transport. The curves correspond to $R_{+}/\omega=2$, $R_{-}/\omega=0.2$, and $\Gamma/\omega=0.5$ (brown dashed), $1$ (blue dash-dotted), and $2$ (red solid). The horizontal gray dash-dotted line indicates zero current, while the colored dotted lines denote the asymptotic slow-switching limit.}
	\label{fig:velocity_comparison}
\end{figure}

Interestingly, this interpretation extends beyond the slow-switching regime for the stochastic protocol. Since $\Omega_1\Omega_2^2\ge\Gamma\bar R^{\,2}\ge\Gamma(\delta R)^2$, the denominator of Eq.~(\ref{v_final_random}) is always positive. Therefore, the sign of the current is entirely determined by $-\delta R$, irrespective of the switching rate $\gamma$, as illustrated in Fig.~\ref{fig:velocity_comparison}(b). By contrast, the periodic protocol exhibits a qualitatively richer behavior. In this case, the current direction is determined by the product $\kappa\,\delta R$, so that the Zeno interpretation applies only while $\kappa<0$. Since $\kappa\to-1$ in the slow-switching limit, whereas Appendix~\ref{AppA} shows that $\kappa\sim CT^2$, with $C>0$, as $T\to0^+$, the transport direction changes as the switching period decreases. As illustrated in Fig.~\ref{fig:velocity_comparison}(a), the current reverses direction, changing from the direction favored by the lattice configuration with the smaller measurement rate to that favored by the one with the larger measurement rate. Similar current reversals have been reported in a variety of directed-transport models~\cite{ReimannR,CasadoPascual2006,SalgadoGarcia2006}. In the present model, the current reversal reflects the crossover from a Zeno- to an anti-Zeno-dominated transport regime.

\begin{figure}[t]
	\centering
	\includegraphics[width=\linewidth]{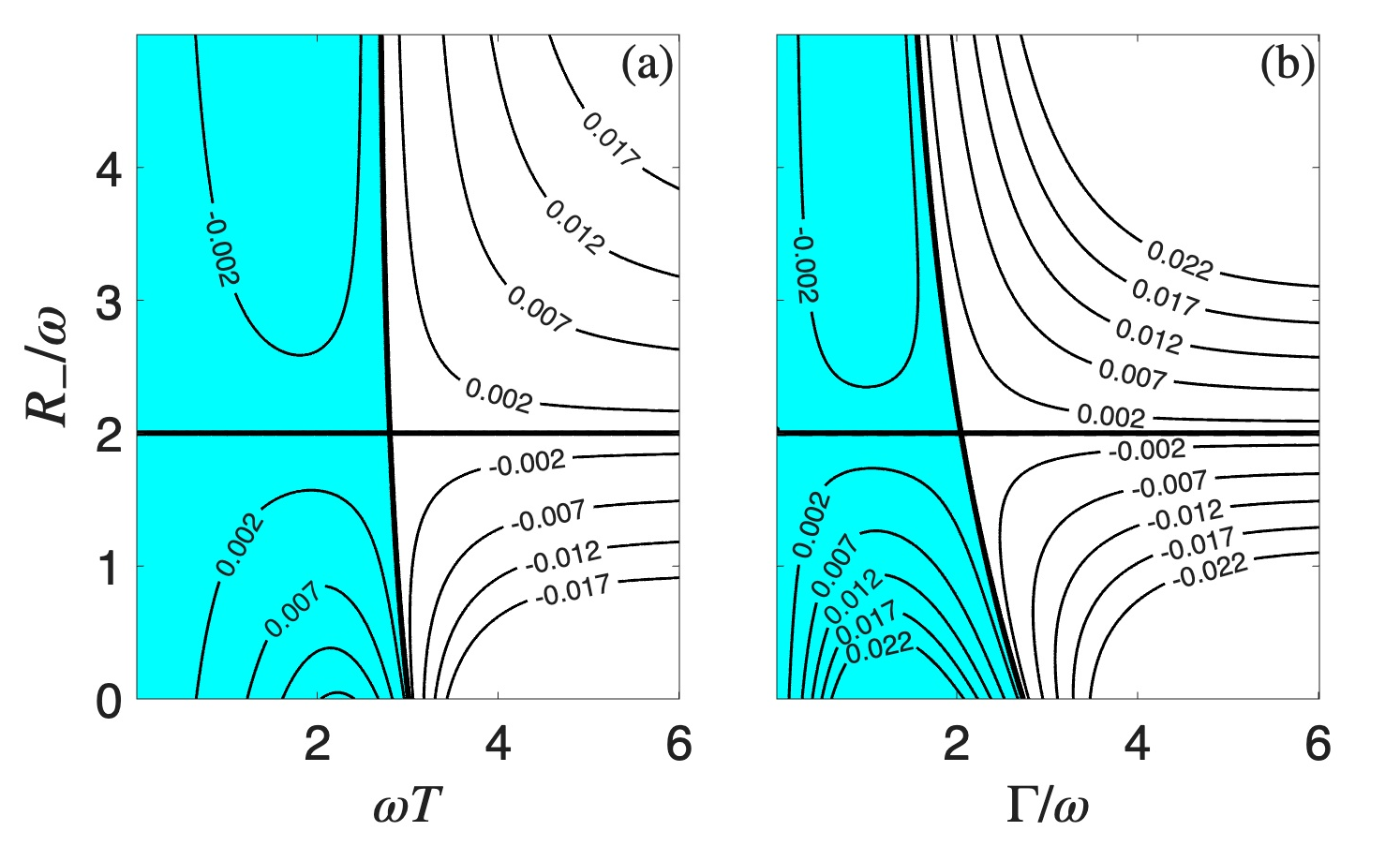}
	\caption{Transport-regime diagram for the periodic protocol. Black curves are contour lines of the dimensionless long-time averaged velocity $v_\infty/(\omega L)$. The thick solid curves indicate the zero-current boundaries ($v_\infty=0$). The cyan-shaded region denotes the anti-Zeno-dominated transport regime ($\kappa>0$), while the unshaded region corresponds to the Zeno-dominated regime ($\kappa<0$), with the two regimes separated by the boundary $\kappa=0$. (a) Dependence on $\omega T$ and $R_-/\omega$ for $\Gamma/\omega=0.5$. (b) Dependence on $\Gamma/\omega$ and $R_-/\omega$ for $\omega T=2$. In both panels, $R_+/\omega=2$.}
	\label{fig:contour_curves}
\end{figure}

\begin{figure}[t]
	\centering
	\includegraphics[width=\linewidth]{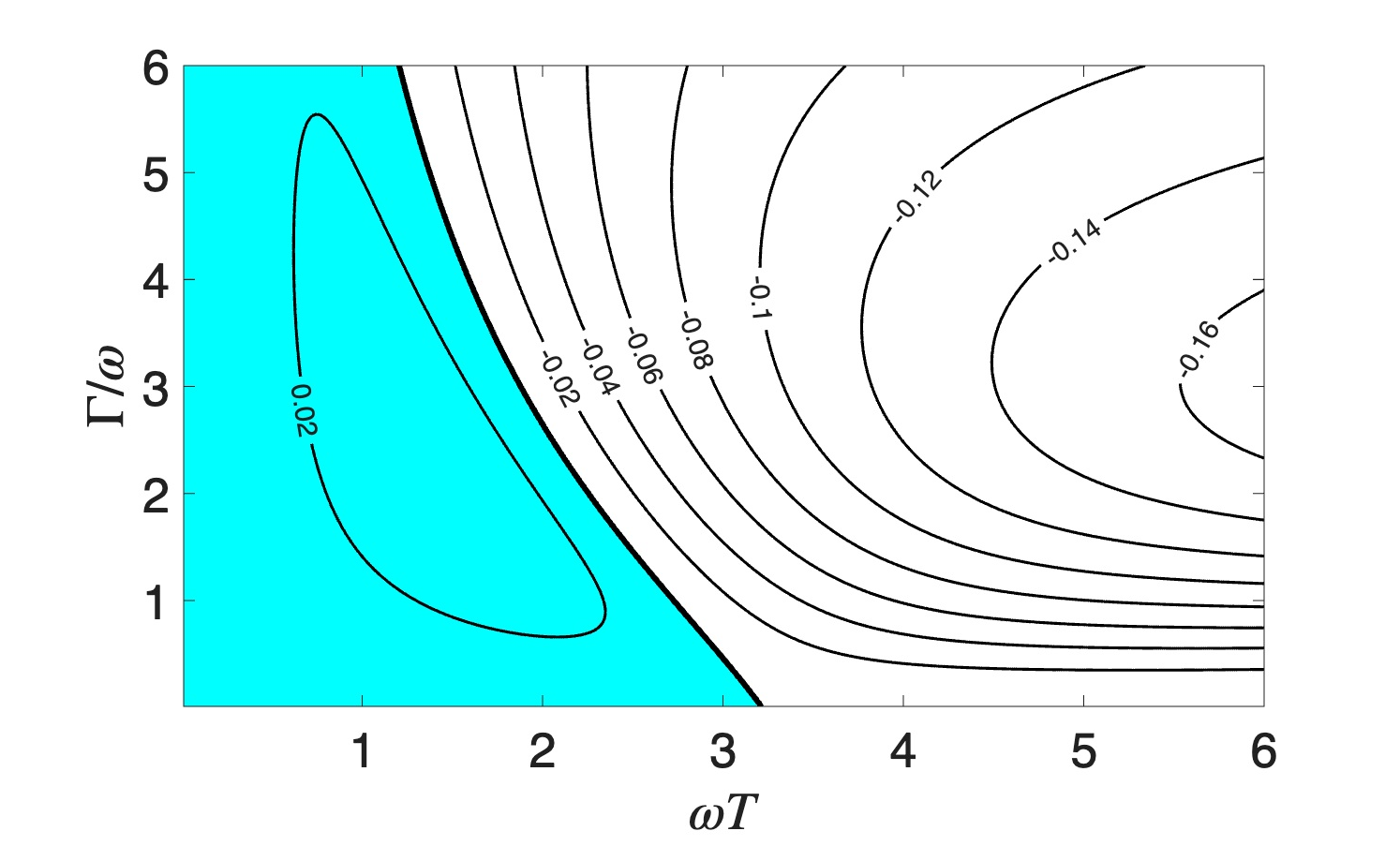}
	\caption{Transport-regime diagram in the $(\omega T,\Gamma/\omega)$ plane for the periodic protocol, corresponding to the parameters $R_+/\omega=2$ and $R_-/\omega=0.2$. Black curves are contour lines of the dimensionless long-time averaged velocity $v_\infty/(\omega L)$. The thick solid curve indicates the zero-current boundary ($v_\infty=0$). The cyan-shaded region denotes the anti-Zeno-dominated transport regime ($\kappa>0$), while the unshaded region corresponds to the Zeno-dominated regime ($\kappa<0$); the boundary is defined by $\kappa=0$.}
	\label{fig:contour_TGamma}
\end{figure}

Figures~\ref{fig:contour_curves} and~\ref{fig:contour_TGamma} summarize the transport regimes predicted by the exact solution~(\ref{velocity1}) for the periodic protocol. Besides the current-reversal boundary at $R_-=R_+$ ($\delta R=0$), where the measurement rates coincide, the transport-regime diagram exhibits the nontrivial boundary $\kappa=0$, which separates the Zeno- and anti-Zeno-dominated regimes. Figure~\ref{fig:contour_curves} shows how this boundary evolves with the modulation period, the measurement rates, and the incoherent hopping rate. Moreover, the anti-Zeno region becomes progressively narrower as $R_-$ increases, the effect being considerably more pronounced in Fig.~\ref{fig:contour_curves}(b) than in Fig.~\ref{fig:contour_curves}(a). Figure~\ref{fig:contour_TGamma} complements this picture by displaying the transport-regime diagram in the $(\omega T,\Gamma/\omega)$ plane for fixed measurement rates, thereby providing a compact view of how the modulation period and incoherent hopping jointly shape the crossover between the two transport regimes.

\paragraph*{Conclusions---\hspace{-10pt}}  We have demonstrated that temporal modulation of the measurement rate provides a mechanism for controlling stationary transport in open quantum systems exclusively through measurement backaction. Using a minimal exactly solvable model, we obtained exact analytical expressions for the long-time current under both periodic and stochastic switching protocols, revealing a common slow-switching limit while showing that only periodic modulation gives rise to a measurement-induced crossover between Zeno- and anti-Zeno-dominated transport regimes accompanied by current reversal. These results establish that the temporal profile of the measurement rate can play the same role as more conventional control parameters, allowing stationary properties to be controlled without modifying the Hamiltonian, the dissipative dynamics, or the measured observable. More generally, they establish measurement-rate modulation as a distinct route for engineering stationary states in continuously monitored quantum systems.

\paragraph*{Acknowledgments---\hspace{-10pt}} This work is part of the project PID2022-136228NB-C22, funded by MICIU/AEI/ 10.13039/501100011033 and by ERDF/EU. This work was also co-financed by the European Union, Ministerio de Hacienda y Funci\'on P\'ublica, FEDER, and Junta de Andaluc\'{\i}a through project SOL2024-31833.

%The authors acknowledge grant PID2022-136228NB-C22 funded by MCIN/AEI/ 10.13039/501100011033 and by ``ERDF A way of making Europe''.   It has also been co-financed by the EU, Ministerio de Hacienda y Funci\'on P\'ublica, FEDER, and Junta de Andaluc\'{\i}a (project SOL2024-31833).

% Produces the bibliography via BibTeX.

\begin{comment}
\clearpage
\onecolumngrid

\begin{center}
	\large\bfseries End Matter
\end{center}

\twocolumngrid
\end{comment}

\appendix

\refstepcounter{section}
\label{AppA}
\paragraph{Appendix A: Details of the Periodic Switching Protocol---\hspace{-10pt}} The piecewise-constant structure of Eqs.~(\ref{ydot}) and (\ref{zdot}) allows the solution to be constructed by concatenating the solutions over the two half periods and imposing continuity at the switching times. Owing to dissipation, the resulting solutions converge, independently of the initial conditions, to a unique asymptotic state that is periodic with period $T$, with the corresponding periodic solutions denoted by $y_{\mathrm{p}}(t)$ and $z_{\mathrm{p}}(t)$. Since the instantaneous velocity given by Eq.~(\ref{avvelo}) is then periodic, the long-time average in Eq.~(\ref{LTZAV}) reduces to its average over one period. Hence,
\begin{equation}
	v_{\infty}
	=
	\frac{L}{2}
	\left[
	2\omega \left(\bar{y}_{\mathrm{p},+}
	+
	\bar{y}_{\mathrm{p},-}\right)
	-
	\Gamma \left(\bar{z}_{\mathrm{p},+}
	+
	\bar{z}_{\mathrm{p},-}\right)
	\right],
	\label{vinf1}
\end{equation}
where
$\bar{y}_{\mathrm{p},\pm}
=
T^{-1}\!\int_{\pm}y_{\mathrm{p}}(t)\,dt$
and
$\bar{z}_{\mathrm{p},\pm}
=
T^{-1}\!\int_{\pm}z_{\mathrm{p}}(t)\,dt$,
the subscripts in $\int_{+}$ and $\int_{-}$ indicating integration over the intervals $[0,T/2)$ and $[T/2,T)$, respectively.

To evaluate $\bar{y}_{\mathrm{p},\pm}$ and $\bar{z}_{\mathrm{p},\pm}$, we integrate Eqs.~(\ref{ydot}) and (\ref{zdot}) over the two intervals introduced above. Using the periodicity conditions $y_{\mathrm{p}}(T)=y_{\mathrm{p}}(0)$ and $z_{\mathrm{p}}(T)=z_{\mathrm{p}}(0)$, the resulting equations can be written as the linear algebraic system
\begin{align}
	\pm \Delta y_{\mathrm{p}}
	&=
	-\frac{T}{2}
	\left(
	\Gamma+2R_{\pm}
	\right)
	\bar{y}_{\mathrm{p},\pm}
	+
	2\omega T \bar{z}_{\mathrm{p},\pm},
	\label{yppm}\\
	\pm \Delta z_{\mathrm{p}}
	&=
	-2\omega T \bar{y}_{\mathrm{p},\pm}
	-
	\Gamma T
	\left(
	\bar{z}_{\mathrm{p},\pm}
	\mp
	\frac{1}{2}
	\right).
	\label{zppm}
\end{align}
Solving Eqs.~(\ref{yppm}) and (\ref{zppm}) for
$\bar{y}_{\mathrm{p},\pm}$ and
$\bar{z}_{\mathrm{p},\pm}$,
and substituting the resulting expressions into Eq.~(\ref{vinf1}), yields Eq.~(\ref{velocity1}).

Equation~(\ref{velocity1}) still requires the determination of $\Delta y_{\mathrm{p}}$ and $\Delta z_{\mathrm{p}}$, which enter the definition of $\kappa$. To compute these quantities, Eqs.~(\ref{ydot}) and (\ref{zdot}) are solved explicitly within each interval of constant coefficients. Introducing the vector
$\mathbf{Q}(t)=(y_{\mathrm{p}}(t),z_{\mathrm{p}}(t))^{\top}$,
together with the matrices
\begin{equation}
	\mathbf{A}_{\pm}
	=
	\begin{pmatrix}
		-\dfrac{\Gamma}{2}-R_{\pm} & 2\omega \\[1.2ex]
		-2\omega & -\Gamma
	\end{pmatrix},
	\label{Apm}
\end{equation}
their inverses $\mathbf{A}_\pm^{-1}$, the propagators
$\mathbf{\Phi}_{\pm}=e^{\mathbf{A}_{\pm}T/2}$,
and the vector $\mathbf{e}_2=(0,1)^{\top}$,
the solution after each half period can be written as
\begin{align}
	\mathbf{Q}(T/2)
	&=
	\mathbf{\Phi}_+\left[\mathbf{Q}(0)+\Gamma \mathbf{A}_+^{-1}\mathbf{e}_2\right]
	-\Gamma \mathbf{A}_+^{-1}\mathbf{e}_2,
	\label{QT2}
	\\
	\mathbf{Q}(T)
	&=
	\mathbf{\Phi}_-\left[\mathbf{Q}(T/2)-\Gamma \mathbf{A}_-^{-1}\mathbf{e}_2\right]
	+\Gamma \mathbf{A}_-^{-1}\mathbf{e}_2.
	\label{QT}
\end{align}

By imposing the periodicity condition $\mathbf{Q}(T)=\mathbf{Q}(0)$, Eqs.~(\ref{QT2}) and (\ref{QT}) yield a linear system for $\mathbf{Q}(0)$ and $\mathbf{Q}(T/2)$. Solving this system and using
$\mathbf{Q}(T/2)-\mathbf{Q}(0)
=(\Delta y_{\mathrm{p}},\Delta z_{\mathrm{p}})^{\top}$,
one obtains $\Delta y_{\mathrm{p}}$ and $\Delta z_{\mathrm{p}}$. Substituting these expressions into Eq.~(\ref{kappa}) yields
\begin{equation}
	\kappa=\frac{\mathbf{a}^{\top}\left(\mathbf{M}_+\mathbf{N}_++\mathbf{M}_-\mathbf{N}_-\right)\mathbf{e}_2}{\omega T}-1,
	\label{kappa2}
\end{equation}
where $\mathbf{a}=(\Gamma,2\omega)^{\top}$,
\begin{equation}
	\begin{split}
		\mathbf{M}_\pm&=(\mathbf{I}-\mathbf{\Phi}_\pm\mathbf{\Phi}_\mp)^{-1}\\
		&=
		\frac{\left[1-\mathrm{Tr}(\mathbf{\Phi}_+\mathbf{\Phi}_-)\right]\mathbf{I}
			+\mathbf{\Phi}_\pm\mathbf{\Phi}_\mp}
		{1-\mathrm{Tr}(\mathbf{\Phi}_+\mathbf{\Phi}_-)
			+\det(\mathbf{\Phi}_+)\det(\mathbf{\Phi}_-)},
	\end{split}
\end{equation}
and $\mathbf{N}_\pm=
\mathbf{\Phi}_\pm(\mathbf{I}-\mathbf{\Phi}_\mp)\mathbf{A}_\mp^{-1}
-(\mathbf{I}-\mathbf{\Phi}_\pm)\mathbf{A}_\pm^{-1}$,
with $\mathbf{I}$ denoting the $2\times2$ identity matrix. For dissipative dynamics, the matrices $\mathbf{I}-\mathbf{\Phi}_+\mathbf{\Phi}_-$ and $\mathbf{I}-\mathbf{\Phi}_-\mathbf{\Phi}_+$ are invertible, so that $\mathbf{M}_\pm$ are well defined.

Equation~(\ref{kappa2}) admits particularly simple expressions in two asymptotic regimes. In the slow-driving limit, where the period $T$ is much longer than all intrinsic timescales of the system ($T\to\infty$), dissipation causes the propagators $\mathbf{\Phi}_{\pm}$ to vanish exponentially. As a result, the numerator of Eq.~(\ref{kappa2}) approaches a constant, whereas its denominator grows linearly with $T$. Hence, the fraction vanishes as $T\to\infty$, yielding $\kappa\to -1$. In the opposite, fast-driving limit ($T\to0^{+}$), the propagators admit a power-series expansion in $T$. Substituting these expansions into Eq.~(\ref{kappa2}) yields
\begin{equation}
	\kappa \sim
	\frac{\left[8\omega^2+\Gamma \left(\Gamma+2R_{+}\right)\right]
		\left[8\omega^2+\Gamma\left(\Gamma+2R_{-}\right)\right]T^2}
	{96\left[8\omega^2+\Gamma\left(\Gamma+2\bar{R}\right)\right]}.
	\label{kappaasymptoticT0}
\end{equation}

\refstepcounter{section}
\label{AppB}

\paragraph*{Appendix B: Details of the Random Switching Protocol---\hspace{-10pt}}

Let $\langle\cdots\rangle_{\mathrm{st}}$ denote stationary ensemble averages. By the Birkhoff ergodic theorem, the long-time averaged velocity can be written as
\begin{equation}
	v_{\infty}
	=
	\frac{L}{2}
	\left(
	2\omega\langle y\rangle_{\mathrm{st}}
	-
	\Gamma\langle z\rangle_{\mathrm{st}}
	\right),
	\label{v_ergodic}
\end{equation}
where $\langle s\rangle_{\mathrm{st}}=0$ follows from the symmetric switching rates.

Taking stationary averages of Eqs.~(\ref{ydot}) and (\ref{zdot}), and writing
$R_{s(t)}=\bar{R}+\delta R\,s(t)$,
one obtains
\begin{align}
	\left(
	\frac{\Gamma}{2}
	+
	\bar{R}
	\right)
	\langle y\rangle_{\mathrm{st}}
	+
	\delta R\,
	\langle sy\rangle_{\mathrm{st}}
	-
	2\omega
	\langle z\rangle_{\mathrm{st}}
	&=
	0,
	\label{st1}
	\\[0.8ex]
	2\omega
	\langle y\rangle_{\mathrm{st}}
	+
	\Gamma
	\langle z\rangle_{\mathrm{st}}
	&=
	0.
	\label{st2}
\end{align}

To close the system, we multiply Eqs.~(\ref{ydot}) and (\ref{zdot}) by $s(t)$ and take stationary averages. The resulting equations involve the mixed moments $\langle s\dot{y}\rangle_{\mathrm{st}}$ and $\langle s\dot{z}\rangle_{\mathrm{st}}$, which are evaluated using the Shapiro--Loginov formula~\cite{shapiro1978,bena2006}. For a dichotomous Markov process,
$
\frac{d}{dt}\langle sy\rangle
=
\langle s\dot{y}\rangle
-
2\gamma\langle sy\rangle,
$
and analogously for $z$. Since stationary averages are time independent,
$
\langle s\dot{y}\rangle_{\mathrm{st}}
=
2\gamma\langle sy\rangle_{\mathrm{st}}$ and
$\langle s\dot{z}\rangle_{\mathrm{st}}
=
2\gamma\langle sz\rangle_{\mathrm{st}}.
$
This yields
\begin{align}
	\delta R\,
	\langle y\rangle_{\mathrm{st}}
	+
	\left(
	\frac{\Gamma}{2}
	+
	\bar{R}
	+
	2\gamma
	\right)
	\langle sy\rangle_{\mathrm{st}}
	-
	2\omega
	\langle sz\rangle_{\mathrm{st}}
	&=
	0,
	\label{st3}
	\\[0.8ex]
	2\omega
	\langle sy\rangle_{\mathrm{st}}
	+
	(\Gamma+2\gamma)
	\langle sz\rangle_{\mathrm{st}}
	&=
	\Gamma.
	\label{st4}
\end{align}
Solving the linear system formed by Eqs.~(\ref{st1})--(\ref{st4}) and substituting the resulting expressions for $\langle y\rangle_{\mathrm{st}}$ and $\langle z\rangle_{\mathrm{st}}$ into Eq.~(\ref{v_ergodic}) yields Eq.~(\ref{v_final_random}).

%\bibliography{RefQRW} 
\providecommand{\noopsort}[1]{}\providecommand{\singleletter}[1]{#1}%

\end{document}